\documentclass[twocolumn]{aastex62}

\pdfoutput=1

\usepackage [english]{babel}
\usepackage [autostyle, english = american]{csquotes}
\MakeOuterQuote{"}
\usepackage{mathtools}
\usepackage{graphicx}
\usepackage{footmisc}
\usepackage{bm}
\usepackage[T1]{fontenc}

\graphicspath{{./}{figures/}}

\received{December 24, 2018}
\revised{March 10, 2019}
\accepted{March 11, 2019}

\submitjournal{AJ}

\turnoffedit
\begin{document}

\title{Three direct imaging epochs could constrain the orbit of Earth 2.0 inside the habitable zone}

\correspondingauthor{Claire Marie Guimond}
\email{claire.guimond2@mail.mcgill.ca}

\author[0000-0003-1521-5461]{Claire Marie Guimond}
\affiliation{Department of Earth \& Planetary Sciences, McGill University, 3450 rue University, Montr\'eal, QC Canada, H3A 0E8}

\author{Nicolas B. Cowan}
\affiliation{Department of Earth \& Planetary Sciences, McGill University, 3450 rue University, Montr\'eal, QC Canada, H3A 0E8}
\affiliation{Department of Physics, McGill University, 3600 rue University, Montr\'eal, QC Canada H3A 2T8}

\begin{abstract}
Space-based direct imaging missions (HabEx, LUVOIR) would observe reflected light from exoplanets in the habitable zones of Sun-like stars. The ultimate\added{---but not sole---}goal of these concept missions is to characterize such planets. Knowing an exoplanet's orbit would help two-fold: (i) its semi-major axis informs whether the planet might harbour \replaced{liquid surface}{surface liquid} water, making it a priority target; and (ii) predicting the planet's future location would tell us where and when to look. The science yields of HabEx and LUVOIR depend on the number, cadence, and precision of observations required to establish a planet's orbit. We produce mock observations using realistic distributions for the six Keplerian orbital parameters, experimenting with both Beta and uniform eccentricity distributions, and accounting for imperfect astrometry ($\sigma$ = 3.5 mas) and obscuration due to the inner working angle of a high-contrast imaging system (IWA = 31 mas). Using Markov chain Monte Carlo methods, we fit the orbital parameters, and retrieve their average precisions and accuracies as functions of cadence, number of epochs, and distance to target. Given the time at which it was acquired, each image provides two data: the $x$ and $y$ position of the planet with respect to its star. Parameter retrieval based on one or two images is formally under-constrained; yet the semi-major axis posterior can be obtained semi-analytically. For a \replaced{1 AU, 10 pc target}{planet at 1 AU around a star at a distance of 10 pc}, three epochs constrain the semi-major axis to within $\lesssim$5\%, if \replaced{the images are}{each image is} taken at least 90 days apart.
\end{abstract}


\section{Introduction} \label{sec:intro}

\begin{deluxetable*}{p{5cm}clll}
\tablecaption{The Keplerian orbital parameters and their priors. The prior range refers to the allowable parameter space explored in the Markov chain, while truth values are drawn from the input range.\label{tab:params}}
\tablehead{
\colhead{Name} & \colhead{Symbol} & \colhead{Prior distribution} & \colhead{Prior range} & \colhead{Input}}
\tabletypesize{\small}
\startdata
Semi-major axis & $a$ & Uniform in natural log & [0.01, 50] AU & [0.95, 1.70] AU \\
Eccentricity & $e$ & Beta ($\sigma$=0.081) or uniform & [0, 1) & [0, 1) \\
Inclination & $i$ & Uniform in cosine & [0, $\pi$] & [0, $\pi/2$] \\
Argument of periapsis & $\omega_p$ & Uniform   & [0, $2\pi$)  & [0, $2\pi$)\\
Longitude of the ascending node & $\Omega$ & Uniform  & [0, $2\pi$) & [0, $2\pi$) \\
Mean anomaly at first epoch & $M_0$ & Uniform & [0, $2\pi$) & [0, $2\pi$) 
\enddata
\end{deluxetable*}

The first generation of space-based direct imaging missions would observe planets in the habitable zones of Sun-like stars. As recommended by the National Academies of Sciences, Engineering, and Medicine's recent Exoplanet Science Strategy, "NASA should lead a large strategic direct imaging mission capable of measuring the reflected-light spectra of temperate terrestrial planets orbiting Sun-like stars."\footnote{https://www.nap.edu/read/25187/} To acquire these spectra, we must first plan the missions cost-effectively. 

\added{Yet identifying contenders for Earth-like planets is not straightforward with direct imaging alone. }A single direct imaging observation would distinguish Earths from similarly-bright sub-Neptunes with a one-in-five success rate, due to degeneracies between radius, albedo, and phase\deleted{ \citep{Guimond18}}. If the planet's semi-major axis is known, on the other hand, the \replaced{false positive}{success} rate is one in two\added{ \citep{Guimond18}}.


In this paper we help prepare for future direct imaging missions such as HabEx and LUVOIR by determining how many visits are required to each newfound planet before we can confidently know its semi-major axis. In doing so, we hope to quantify a key input parameter in more comprehensive mission simulations \citep[see][]{Stark18}.

\subsection{Images to orbits}

A single epoch of direct imaging provides the two-dimensional position of the planet in the sky plane relative to its host star: $x$ and $y$, where the star is at the origin. Six parameters are needed to uniquely describe the three-dimensional Keplerian orbit of a planet (listed in Table \ref{tab:params}). \replaced{Since E}{Given the time at which it was taken, e}ach image \replaced{provides two data,}{constrains} $x$ and $y$, \added{so }at most three measurements should be required to fit all six parameters.

The fitting methods employed in previous direct imaging orbit retrieval efforts come from either Markov chain Monte Carlo (MCMC) or Bayesian rejection sampling \citep{Pueyo15, DeRosa15, Rameau16, Wang16, Wang18, Blunt17, Kosmo18}. The latter type\replaced{, as demonstrated carefully in \citet{Blunt17}, are designed}{ are carefully demonstrated in \citet{Blunt17}} for closely-spaced observations over tiny orbital coverage ($\sim$3\%). 

\citet{Mede17} present a software package to simultaneously fit direct imaging observations and radial velocity observations. Such an approach may be fruitful for eventual Earth twins if high precision radial velocity can be sensitive to the 10 cm/s signals \citep{Fischer16}.

As for future direct imaging, previous design reference missions for HabEx and LUVOIR have not always implemented an optimized number or cadence of observations. \citet{Stark14} look only at single-visit yields; \citet{Stark16, Stark15} do not consider that a mission would revisit candidates for the strict purpose of establishing their orbits, assuming that most stars would be revisited regardless to increase the total yield.

In this paper, we anticipate space-based observations from \deleted{ WFIRST,} HabEx\deleted{,} and LUVOIR. The work presented here is fundamentally distinct from these earlier efforts in that we explicitly quantify the number, cadence, and precision of observations required to establish, for targets at any distance, the orbit of a planet within its star's habitable zone.

\subsection{A note on habitable zones}

The circumstellar habitable zone (HZ) is a theoretical shell around a star within which planets can harbour liquid water at their surfaces. The inner and outer edges of the HZ have been variously modeled based on assumptions about climatic systems. Early habitable zone estimates \citep{Hart79} were physically driven by the destabilizing ice-albedo and water vapour feedbacks and were narrower than in current conceptions.

The HZ estimates we use come from physics gleaned from the Earth. Assuming the surface of a planet has some exposed silicate rock, CaSiO$_3$ minerals in the rock will react to consume atmospheric CO$_2$. The rate of this reaction increases exponentially with temperature, so the planetary effect of warming the atmosphere is to sequester more carbon and weaken the greenhouse effect \citep{Walker81, Kasting88, KastingToon89}. This is the well-known silicate weathering feedback, and is the basis of the HZs proposed in \citet{Kasting93} and updated in \citet{Kopparapu13}.

This conception of the HZ has yet to be empirically validated \citep[but it could through statistical study of exoplanet properties;][]{Bean17}. If a planet does not have exposed surface rock, for example, then the HZs given by a model based on the silicate weathering feedback might not apply \citep[e.g.][]{Abbot12}. We therefore base our analysis on a variety of "hypothetical" HZ widths\added{, defined as 0.01, 0.05, 0.1, 0.25, and 0.5~AU. For completeness, we also consider the inner and outer HZ limits proposed by \citet{Kopparapu13}}. The semi-major axis precision that would satisfy us depends on the sizes of these HZs relative to our semi-major axis estimate. 

\section{The orbit-retrieval model} 

This work quantifies, statistically, the orbit retrieval accuracy and precision for many simulated HZ planets. We simulate planets with each orbital element randomized\added{ according to the distributions in Table \ref{tab:params}, and assuming one planet per star}. For each epoch of each planet, we calculate the planet's $(x, y)$ position relative to its star, given a fixed cadence in days and with added Gaussian noise\replaced{, $\sigma_\theta$ = 3.5~mas.}{---here we adopt $\sigma_{\theta} = 3.5$~mas as the baseline astrometric precision \citep{HabEx, LUVOIR}.} Then we retrieve the posteriors on orbital elements from our synthetic dataset using MCMC. We repeat this numerical experiment under varying assumptions, described below.

\begin{figure}
\plotone{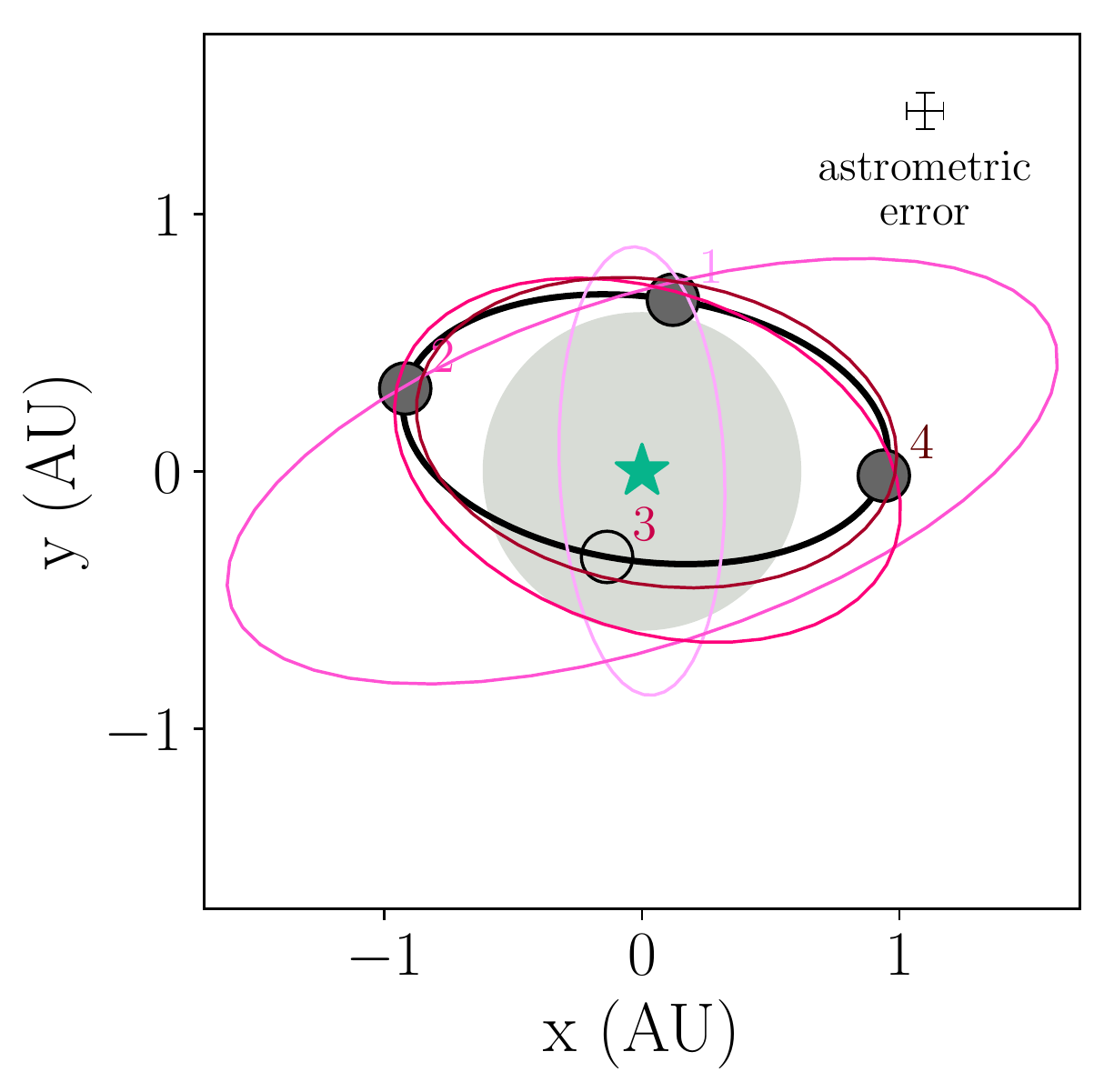}
\caption{\textbf{Demonstration of orbit retrieval.} True orbit (bold black ellipse) shown to scale with the region obscured by the inner working angle (grey circle). The true location of the planet at each epoch is shown as a small circle with radius corresponding to astrometric error; the hollow circle at the third epoch indicates that the planet is inside the inner working angle and not detected. The retrieved orbits after each epoch are shown from light (epoch 1) to dark (epoch 4). For this nominal planet, $r=20$~pc, $\sigma_\theta = 3.5$~mas, and the observations are spaced 90 days apart. 
\label{fig:orbdemo}}
\end{figure}

\paragraph{Eccentricity distribution}Since the eccentricity distribution of HZ terrestrial exoplanets is unconstrained, we repeat our experiment assuming different eccentricity distributions---a beta distribution with parameters $a = 0.867$, $b = 3.03$ \citep{Kipping13, Nielsen08}, or a uniform distribution---for both the underlying true distribution, and the prior distribution used in the retrieval.

\paragraph{Cadence}The retrieval experiments are repeated, using the same synthetic orbits, for cadences of 30, 90, 180, and 270~days. \added{These cadences were chosen as a starting point because they are well-separated round numbers.} This \added{part of the study} reveals which cadence gives the best fit, for the same number of epochs. Our result is optimized for orbital periods within the G2V habitable zone: we expect the best cadence to be set by how much the planet has moved, which is some fraction of the period. In principle, a mission simulation would consider data from previous epochs to choose when to revisit a star. We take the simpler approach of a fixed cadence to obtain a broad result, which could conveniently be applied to mission prognoses.

\paragraph{Distance}The distance of the target star will impact orbit retrieval through astrometric error and the inner working angle (IWA) of starlight suppression, both of which correspond to greater projected separations at greater distances. We take a baseline distance of 10~pc, and repeat our experiment for distances of 5~pc and 20~pc.\\

Each scenario considers \replaced{20}{100}~planets with orbital parameters $a$, $i$, $e$, $\omega_p$, $\Omega$, and $M_0$, sampled from the underlying distributions (Table \ref{tab:params}). These are the prior distributions described in section \ref{sec:priors}---excepting $e$, which we allow to have different underlying and prior distributions, as detailed later. \added{Our sample size of 100 planets is justified in that halving this number gives similar results.}

To retrieve posterior probability distributions of the orbital parameters, we use {\tt{emcee}}, an MCMC ensemble sampler \citep{DFM13}. \replaced{As readers will know, t}{T}he goal of any MCMC implementation is to evaluate the posterior Prob(model|observation, $\sigma$) $\propto$ Prob(observation|model, $\sigma$) $\times$ Prob(model), where $\sigma$ is the measurement precision. Given estimates of the right hand side, we can estimate the left hand side. Thus setting up {\tt emcee} requires:
\begin{enumerate}
\item Estimates of the prior probability distribution, Prob(parameter), for each orbital parameter,
\item A forward model that calculates the planet positions $(x_0, y_0), (x_{1}, y_{1}), ..., (x_{k}, y_{k})$ up to the $k$th epoch given orbital parameters\added{ and $t_0$, $t_1$ ... $t_k$},
\item A likelihood function, Pr(observation|parameters, $\sigma$), that calculates the probability of the observed $(x,y)$ position given the forward model.
\end{enumerate}


The MCMC uses 30 walkers, randomly initialized in a Gaussian ball around the best-fit parmeters from a quick {\tt scipy.optimize} likelihood maximization. The walkers have a burn-in time of 1000 steps and run for up to $5\times10^5$ steps. The rest of the MCMC setup is detailed in subsections \ref{sec:likelihood} through \ref{sec:priors}.

\subsection{Likelihood function for detections}\label{sec:likelihood}

The likelihood of the observed position of the planet at some epoch is a normal distribution centred on the true position with width $\sigma_{xy}$. If we have $k$ images in which the planet is detected, then the log likelihood is
\begin{equation}
\ln L = -\frac{\chi^2}{2} - k(\ln\pi + 2\ln\sigma_{xy}),
\end{equation}
where
\begin{equation}
\chi^2 = \sum^k_{i=1} \frac{(x_{{\rm obv}, i} - x_{{\rm model}, i})^2 + (y_{{\rm obv}, i} - y_{{\rm model}, i})^2}{\sigma_{xy}^2}.
\end{equation}
\deleted{Here we adopt $\sigma_{\theta} = 5$ mas as the baseline astrometric precision \citep{HabEx, LUVOIR}---an error on $x$ of $\sigma_x = \sigma_{xy}/\sqrt{2}$, and an error on $(x, y)$ of $\sigma_{xy} = d\sigma_{\theta}$, where $d$ is the distance to the system.}

\begin{figure}
\plotone{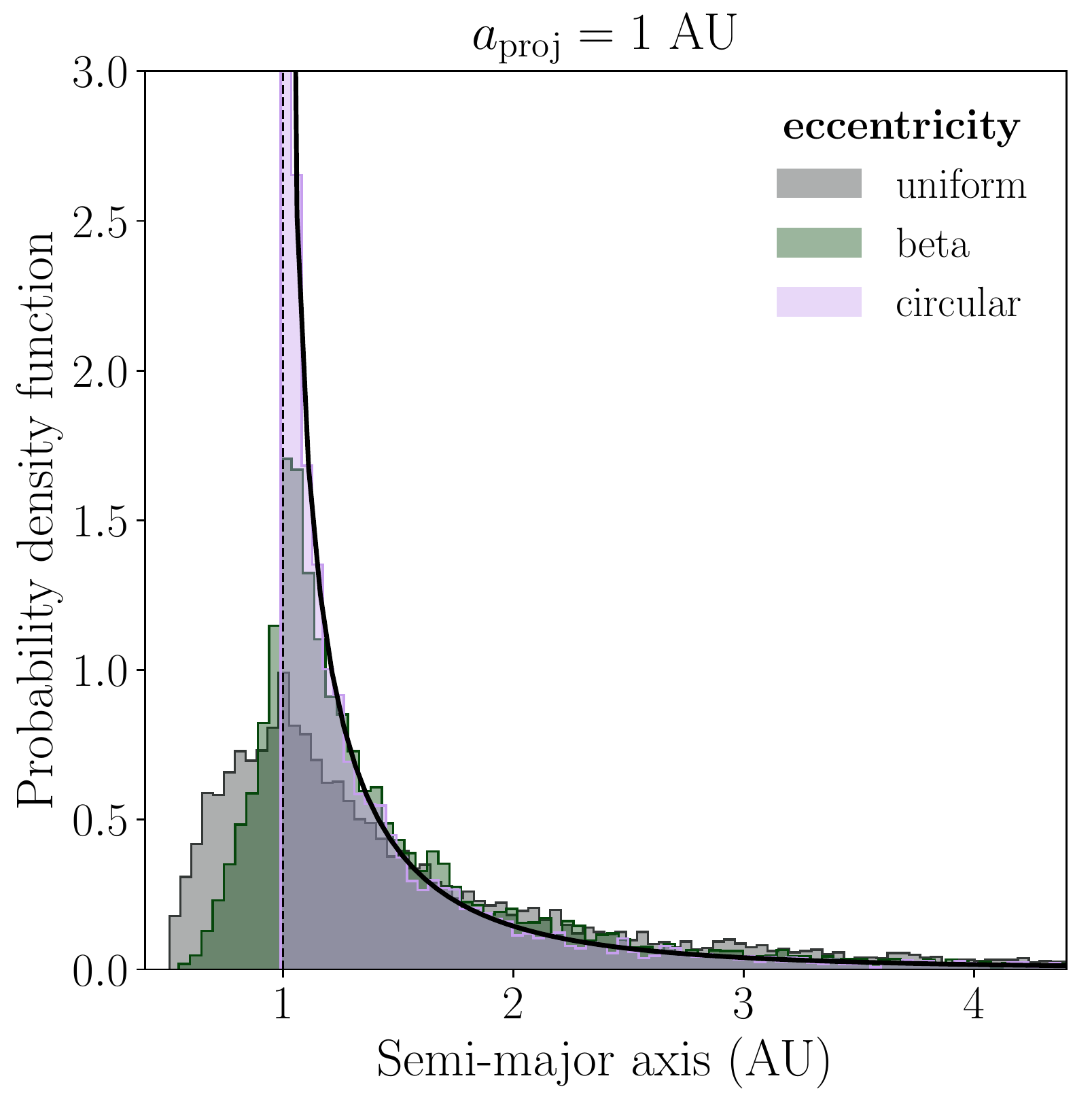}
\caption{\textbf{Probability of semi-major axis given a projected separation.} Normalized histograms of the semi-major axes of $10^3$ simulated planets with $a_{\rm proj}=1$~AU (dashed black line) and random $a$, $i$, $e$, $\omega_p$, $\Omega$, and $M$. Colours indicate different underlying eccentricity distributions: lavender for circular, green for beta \citep{Kipping13,Nielsen08}, and grey for uniform. \added{The thick black line traces the analytic derivation for the circular case (equation \ref{eq:a_dist}).}
\label{fig:a_dist}}
\end{figure}

\subsection{Likelihood function for non-detections}

The first epoch is---by definition---a detection, as we are concerned with the orbit retrieval of actual planets rather than the odds of spotting a planet in the first place. Subsequent epochs may yield non-detections, however. 

A planet may go undetected in an image for two reasons: (i) it is imaged at small projected separation and is occluded along with the starlight, or (ii) it does not reflect enough light. We ignore the latter case: the brightness of the planet depends on the geometric albedo and the scattering phase function, about which we presume nothing. Titan, for example, appears brighter at crescent phase, due to a distinctly non-Lambertian phase function \citep{GM17}. 

The relevant criterion in (i) is the IWA of the telescope, defined as the point where photon transmission through the instrument has decreased by 50\%. We adopt an IWA of 31~mas. \replaced{For a given planet}{For a planet of given size, albedo, and phase function, and given a contrast floor}, there will be a unique IWA at which it becomes undetectable at gibbous phase, and another where it becomes undetectable at crescent phase. We ignore brightness information and hence adopt a single hard-cutoff IWA.

A non-detection therefore still provides a constraint; it means that the angular separation of the planet and the star is less than the IWA. In this way, a nondetection is analogous to a measurement centered at the origin with an uncertainty of $\theta_{\rm IWA}$. \added{We use this information in our retrieval exercises.} If the planet is not detected at some epoch, then any set of parameters that would place it outside the IWA are assigned zero likelihood\added{, Prob(observation|model, $\sigma$) = 0}.

\added{If an orbit-fitting model were incorporating photometry, then a soft-edge IWA would be more appropriate; for example, the non-detection framework in \citet{Ruffio18}. The width of this soft-edge matters when one knows the orbit and is trying to measure the brightness of the planet, which will vary from 0\% to 100\% of its true value across the soft-edge. However, we are interested in the earlier stage, when we do not yet know the orbit, and do not strictly care about the measured brightness of the planet. In other words, there is no need for a soft-edge at this point because the probability of the planet being ``detected'' outside the IWA is always unity}\edit2{, although this statement requires assumptions discussed further in section \ref{sec:limitations}.}

\subsection{Posterior probability for less than three detections}
Given only one or two detections, we expect MCMC to converge slowly, if at all, since the problem is formally under-constrained. \added{The walkers would be exploring a nearly-flat plane of probability; we would not believe the width of their retrieved distribution.} Thus we eschew Kepler's laws for a naive, semi-analytic approach, as long as the number of detections is less than three. 

That is: for a given detection, the posterior probability distribution of $\ln(a)$ is a strong function of the projected separation, $a_{\rm proj}$. For circular orbits we can express this analytically:
\begin{equation}\label{eq:a_dist}
{\rm Prob}(a\mid a_{\rm proj}, e=0) = \frac{1}{\beta^2\sqrt{\beta^2-1}}, \qquad \beta \in [1, \infty),
\end{equation}
where $\beta = a/a_{\rm proj}$. Nonzero eccentricities complicate this analytic distribution, but we can model it numerically for given values of  $a_{\rm proj}$ and $a_{\rm IWA}$ by drawing 1000 random orbits from the priors in Table \ref{tab:params} with $a_{\rm proj}$ fixed at the observed value. \added{We keep drawing all six Keplerian parameters until we have 1000 orbits with the desired $a_{\rm proj}$.} As figure \ref{fig:a_dist} shows, the distribution of semi-major axes of these orbits \replaced{approximates the posterior distribution}{peaks at the true $a$, and is hence a reasonable proxy for the posterior}, for a given measurement of $a_{\rm proj}$. If a planet is not detected at some epoch, we also simulate random orbits, but with $a_{\rm proj}~\in~(0,a_{\rm IWA}$].

For two or more epochs, we construct "independent" posteriors for each $a_{\rm proj}$ observation, and multiply them together. The resulting joint probability distribution usually has one peak. Of course these measurements would not be truly independent, since they are linked by a Keplerian orbit. Thus our semi-analytic posterior distributions give an upper limit on the true uncertainty.

\begin{figure}
\plotone{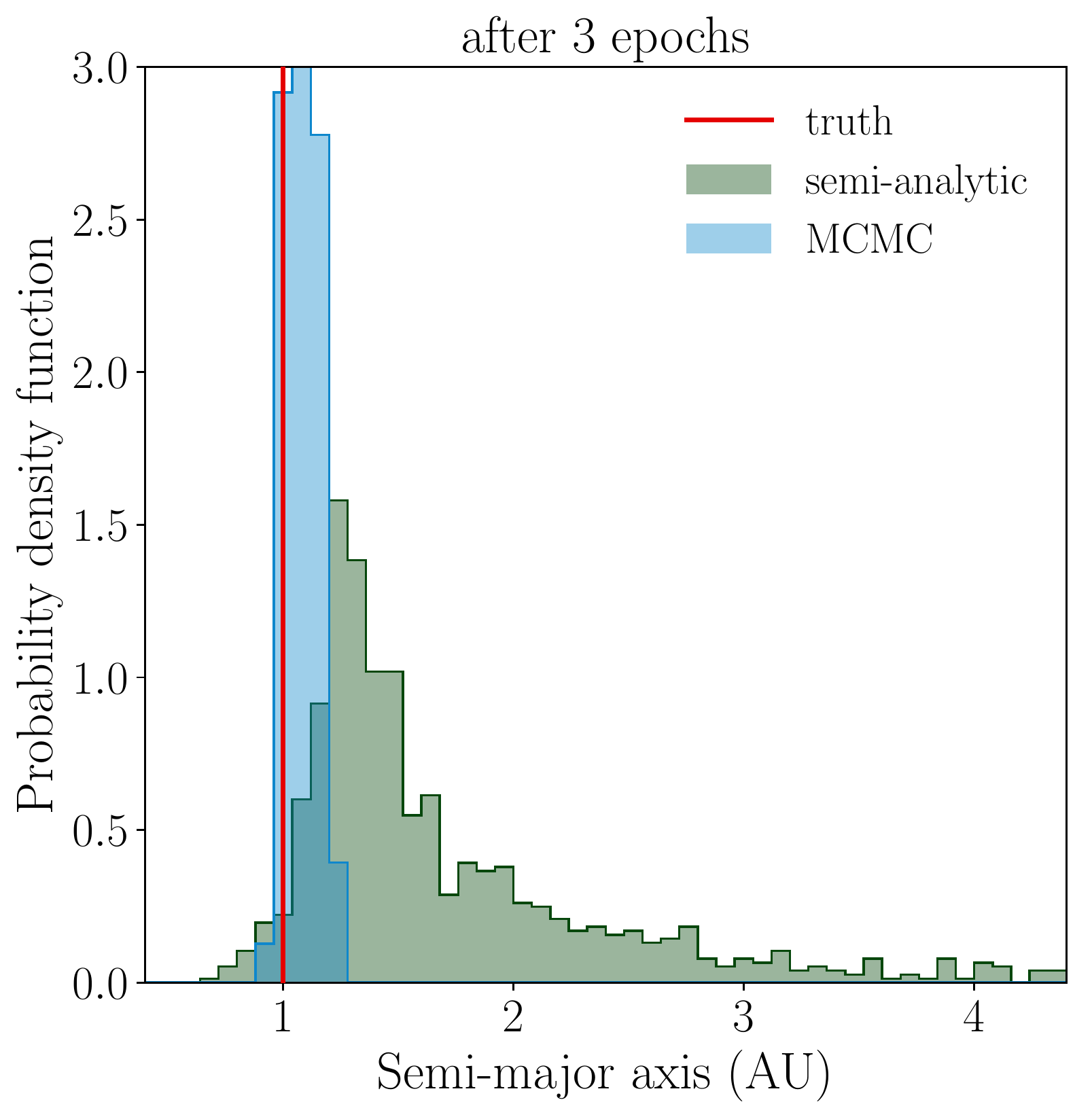}
\caption{\added{\textbf{Comparison of semi-analytic and MCMC posterior retrievals.} \textit{(Green:)} Normalized joint probability density resulting from multiplying together three histograms (figure \ref{fig:a_dist}) for three ``independent'' observations of one planet's $a_{\rm proj}$, assuming 90~days between each image. \textit{(Blue:)} Posterior probability density after three simulated observations of the same system using Markov chain Monte Carlo. The vertical red line shows the true semi-major axis. The improved performance of the MCMC comes from the added use of Keplerian orbits over merely using geometry.} 
\label{fig:a_post_compare}}
\end{figure}

\added{This semi-analytic method is exchanged for a full MCMC once we have three detections, allowing the latter algorithm to converge. Despite the superior speed of the semi-analytic method, it does not use information about the relative timing of the different epochs, and hence does not fully leverage the Keplerian orbit. Figure \ref{fig:a_post_compare} demonstrates the relatively poor performance of the semi-analytic method compared to MCMC after three epochs.}

\subsection{Prior distributions} \label{sec:priors}

As summarized in Table \ref{tab:params}, parameterizations are chosen such that all priors save for eccentricity are flat. The log-uniform prior on semi-major axis is roughly consistent with recent literature \citep{Petigura13, DFM14, Silburt15, Burke15, Christiansen15, Kopparapu18}. We adopt a wide prior range of $a \in [0.1, 50]$~AU.

The mean anomaly at the time of the first epoch is uniformly distributed in $[0, 2\pi)$, as are arguments of periapsis and longitudes of the ascending node. Inclination is uniform in $\cos{i} \in [-1, 1]$.

Eccentricity is drawn from a beta distribution with parameters $a = 0.867, b = 3.03$, as determined for radial velocity planets by \citet{Kipping13} and in agreement with \citet{Nielsen08}. Because this parameter's underlying distribution is especially hard to constrain, we repeat our experiment with a uniform prior and uniformly-distributed true eccentricities, and again with a beta prior and uniformly-distributed true eccentricities (where we would be over-confident).

\begin{figure*}
\epsscale{0.8}
\gridline{\fig{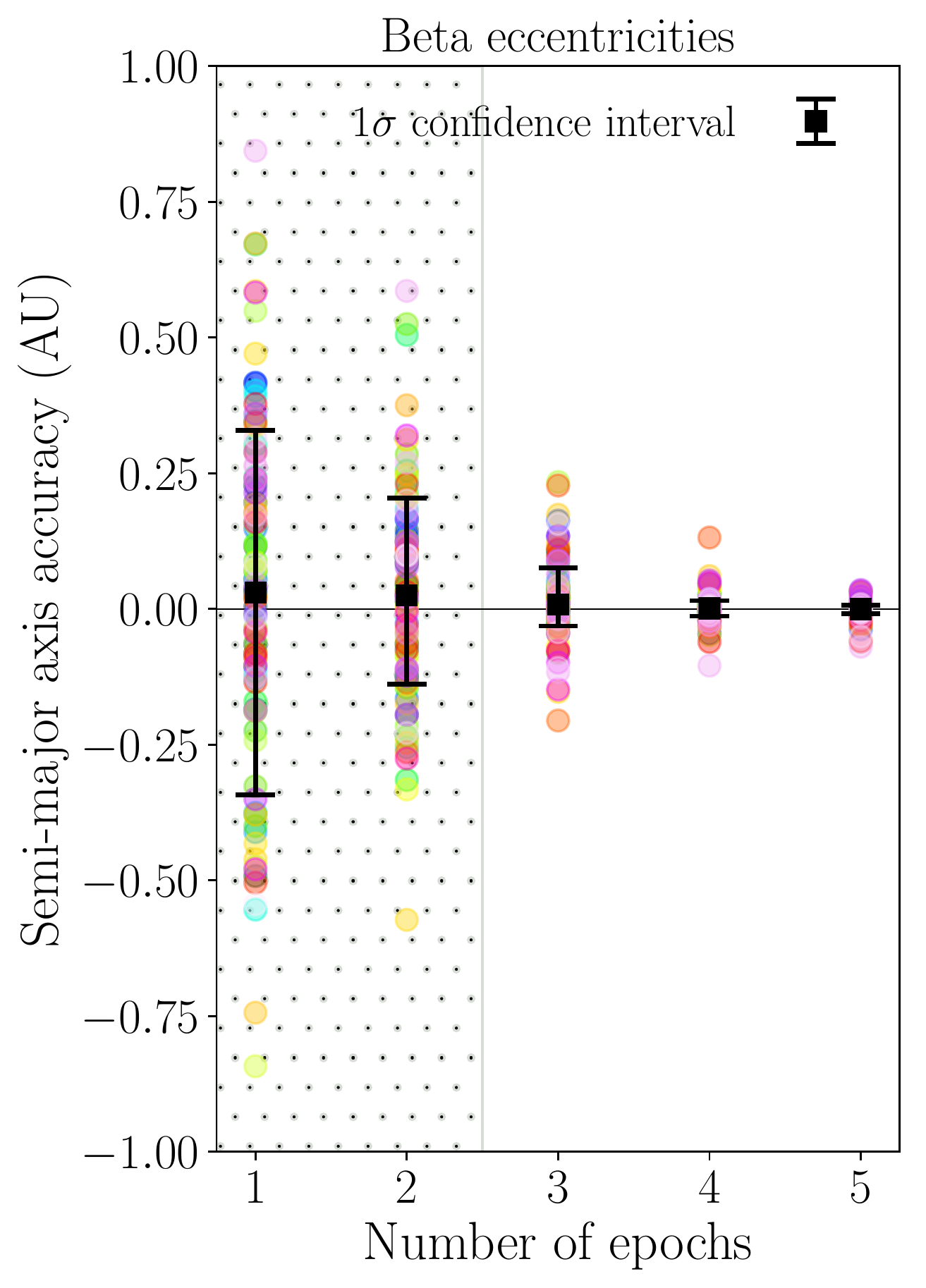}{0.33\textwidth}{(a)}
		  \fig{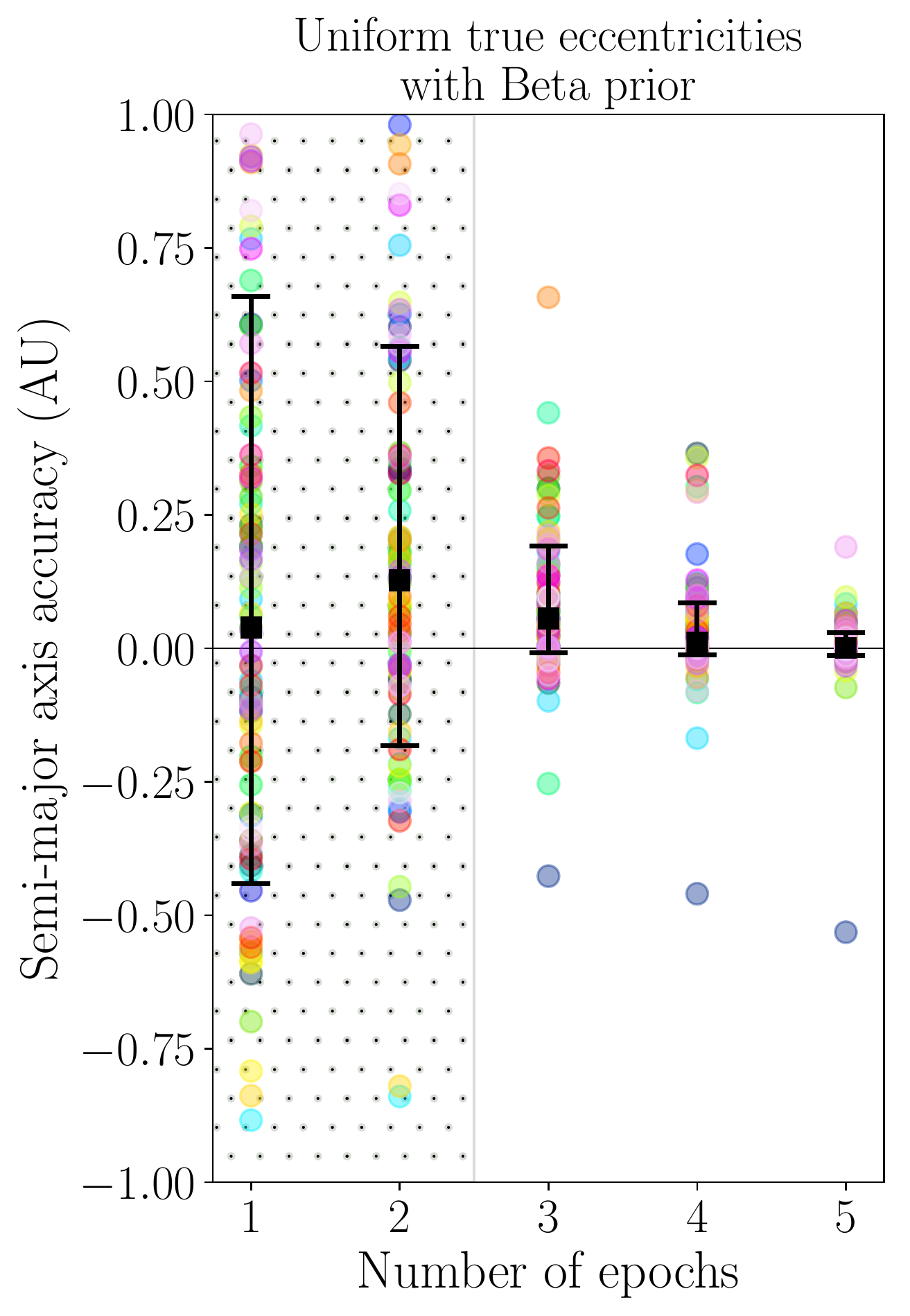}{0.33\textwidth}{(b)}
          \fig{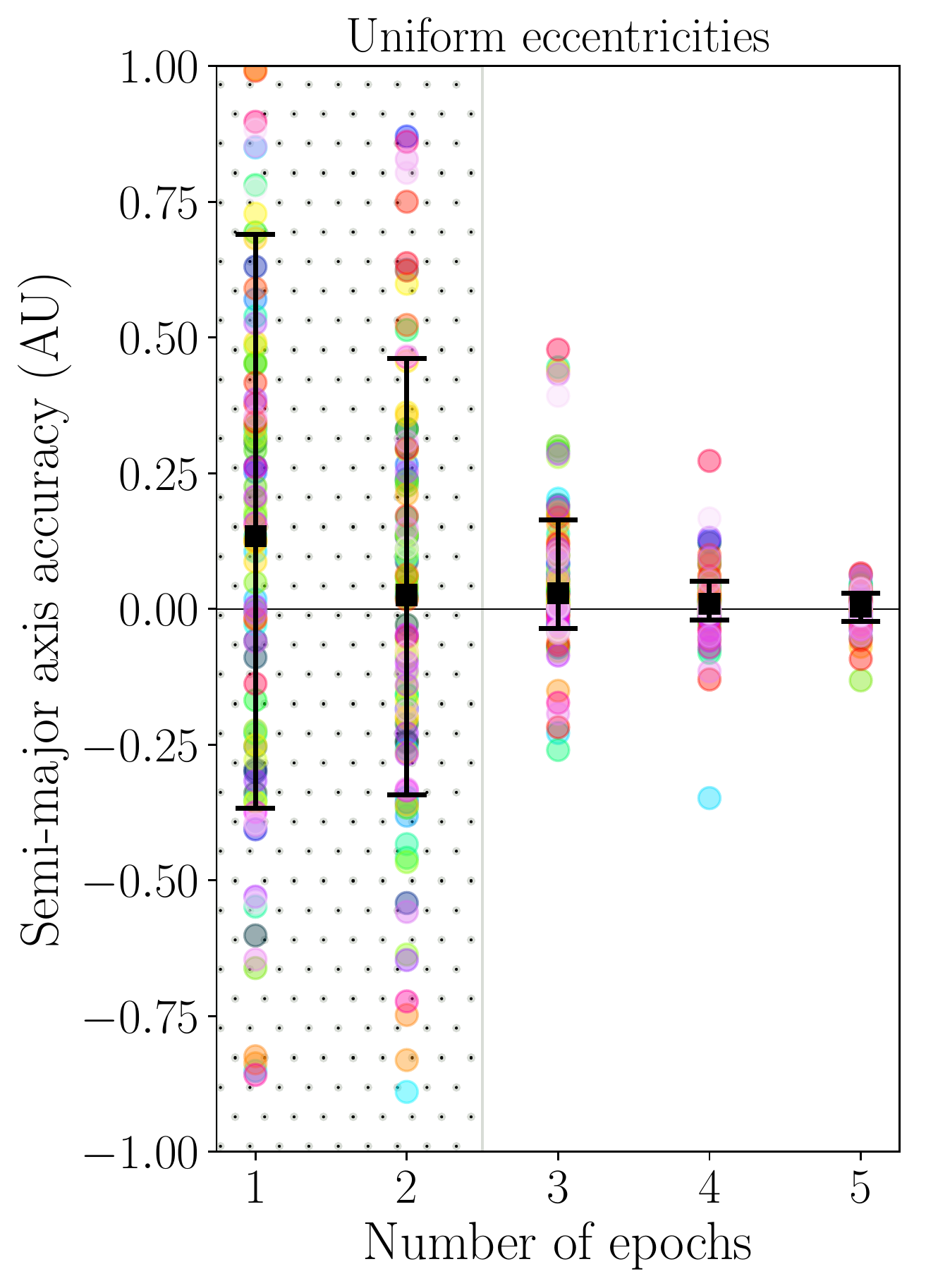}{0.33\textwidth}{(c)}} 
\caption{\textbf{Semi-major axis accuracy by epoch.} Epoch-dependence of accuracy, defined as $(a_{\rm fitted} - a_{\rm true})$, where $a_{\rm fitted}$ is the median of the posterior semi-major axis distribution from MCMC. All simulated planets (\replaced{50}{100} in each scenario) have a distance of 10~pc and are observed every 90 days. Each dot colour represents one planet. The black error bars show the 16th and 84th percentiles of the retrieved accuracy (i.e., the standard error). Until a planet is detected three times, orbits are retrieved with a semi-analytic method as opposed to MCMC and represent a conservative upper limit; the hatched region marks off this distinction. Eccentricities are either: \textit{(a)} drawn from a low-dispersion beta distribution for both the underlying true orbits and the prior; \textit{(b)} drawn from a uniform distribution for the underlying truths and a beta distribution for the prior; or \textit{(c)} drawn from a uniform distribution for both input and prior.
\label{fig:error}}
\end{figure*}

\begin{figure*}
\epsscale{0.8}
\gridline{\fig{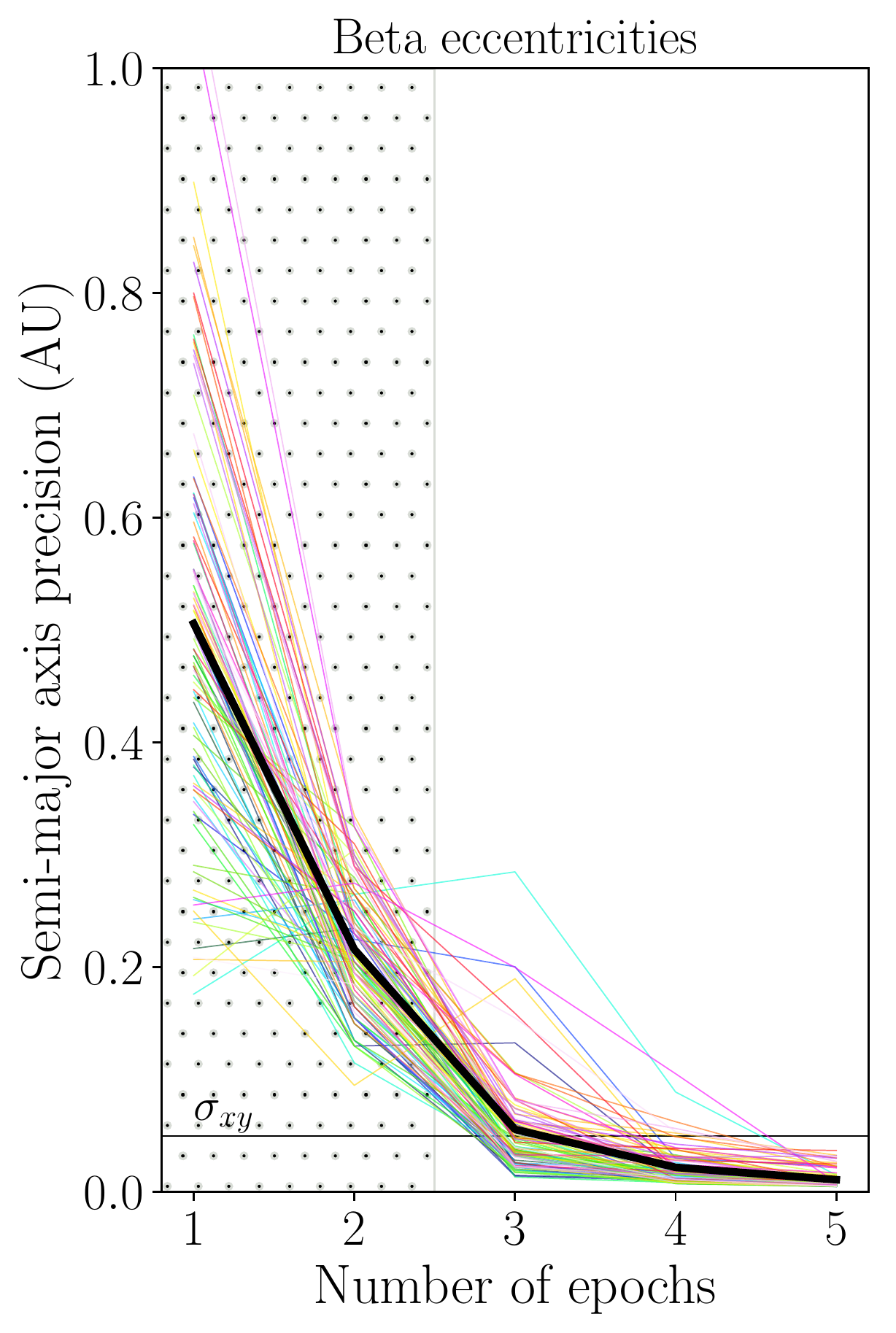}{0.3\textwidth}{(a)}
		  \fig{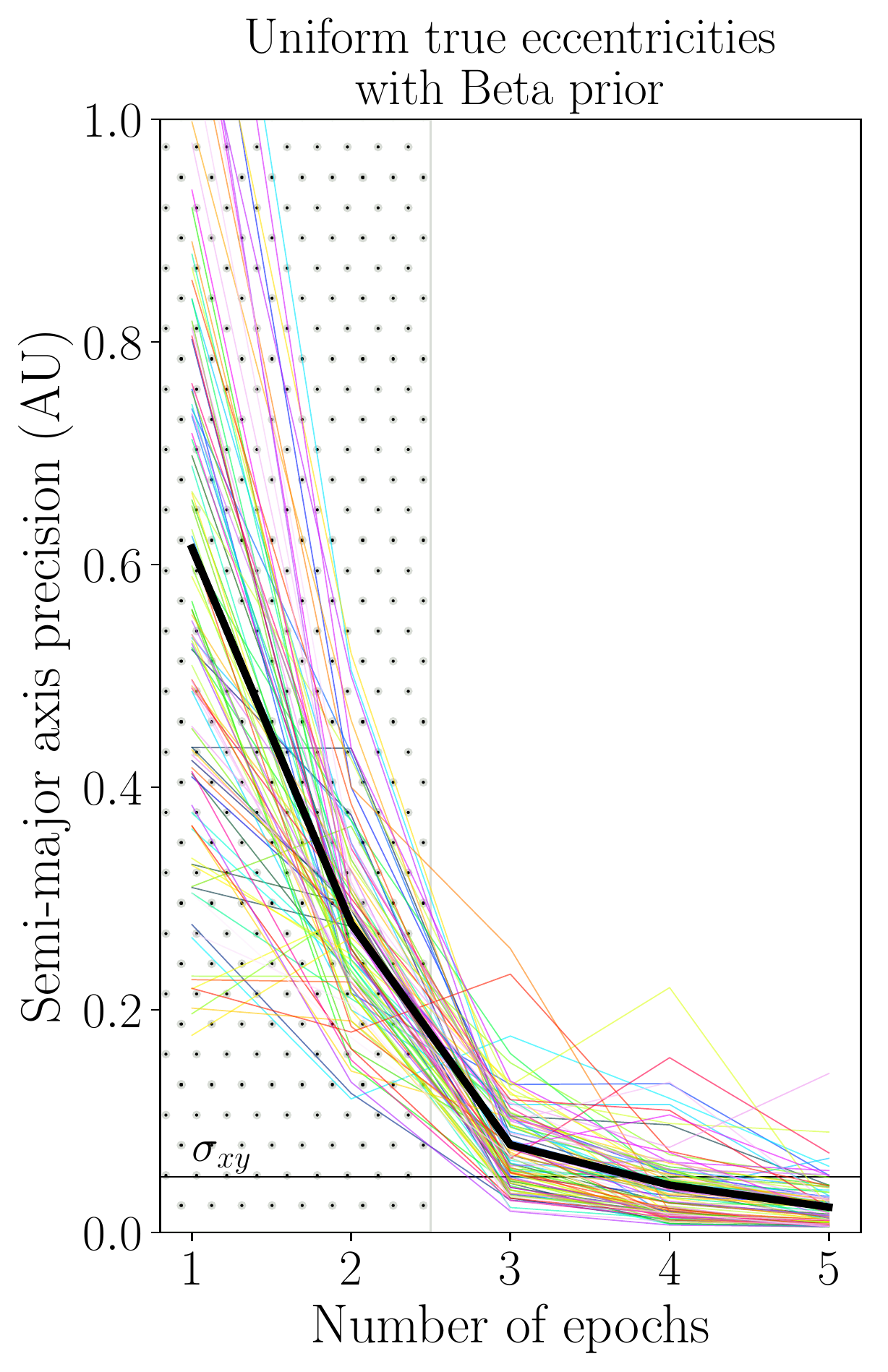}{0.3\textwidth}{(b)}
          \fig{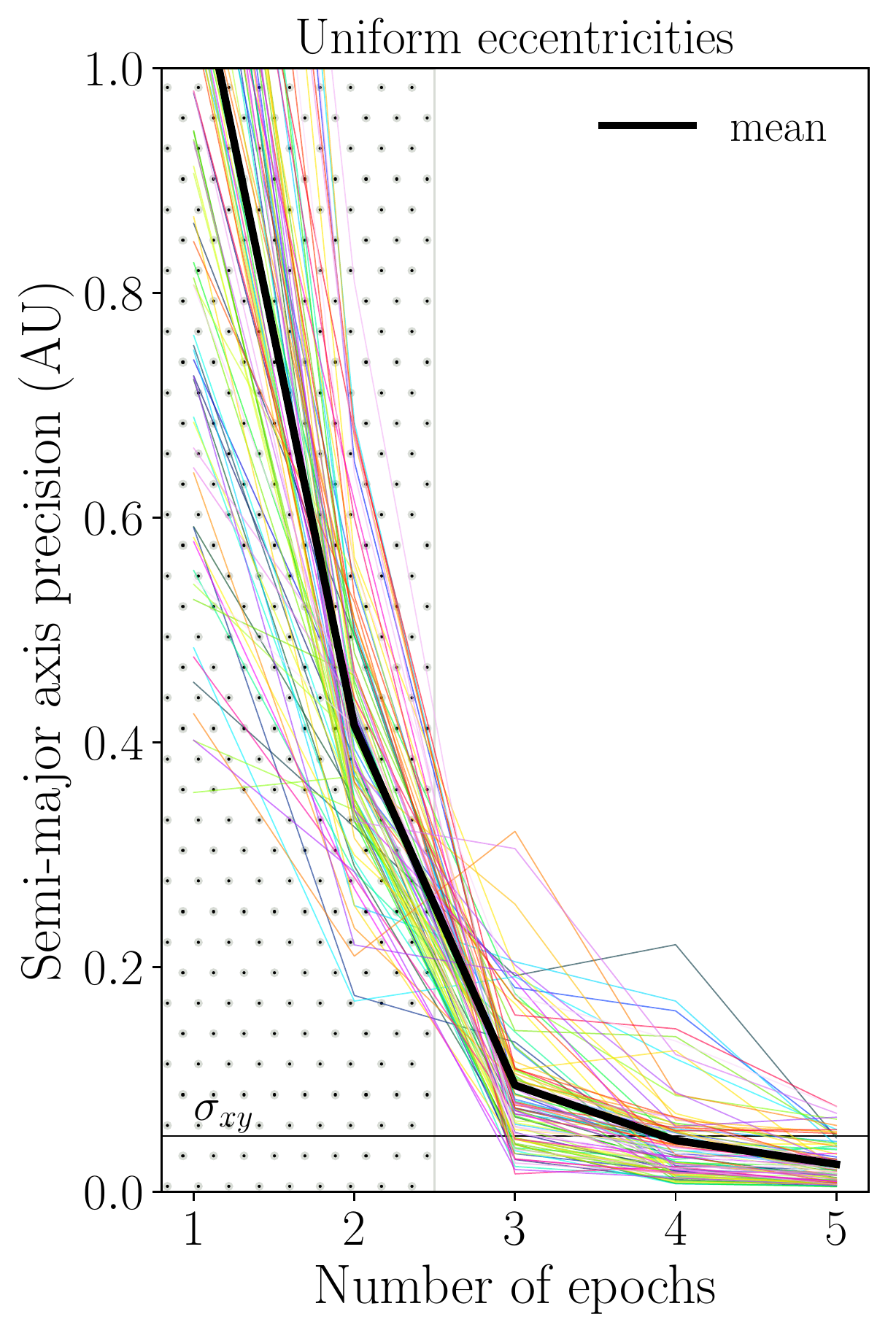}{0.3\textwidth}{(c)}}
\caption{\textbf{Semi-major axis precision by epoch.} 1$\sigma$ precision on semi-major axis retrieval as a function of number of epochs. Coloured lines show the \replaced{50}{100} individual runs using the same scheme as in figure \ref{fig:error}, while the bold black line is the mean precision (the median is slightly better). Precision is defined as $(a_{+1\sigma} - a_{-1\sigma})/2$, where the subscripts refer to the retrieved upper and lower limits of the 68\% confidence interval. The horizontal line indicates the measurement uncertainty for 3.5~mas astrometry at 10~pc. Until a planet is detected three times, fits are done with a semi-analytic method as opposed to MCMC and represent a conservative upper limit; the hatched region marks off this distinction. In this scenario, planets are observed every 90 days, and eccentricities are either: \textit{(a)} drawn from a low-dispersion beta distribution for both the underlying true orbits and the prior; \textit{(b)} drawn from a uniform distribution and the prior is a beta distribution; or \textit{(c)} drawn from a uniform distribution and the prior is uniform. Around the third epoch, the problem becomes constrained, and we see an inflection point in precision.
\label{fig:precision}}
\end{figure*}
\section{Results} \label{sec:results}

\begin{deluxetable}{lrrrrrr}
\tablecaption{Z-score statistics for data in figures \ref{fig:error} and \ref{fig:precision}. Cases \textit{(a)} and \textit{(c)} are fine, but \textit{(b)} is dangerous. \label{tab:zscores}}
\tablehead{
\colhead{} & \multicolumn{2}{c}{\textit{(a)} Beta} & \multicolumn{2}{c}{\textit{(b)} Mixed} & \multicolumn{2}{c}{\textit{(c)} Uniform}}
\tabletypesize{\small}
\startdata
       & mean & stdev. & mean & stdev. & mean & stdev. \\
Epoch 3 & -0.47 & 1.19  & -1.02 & 2.56  & -0.18 & 1.26 \\
Epoch 4 & -0.10 & 0.99 & -0.28 & 2.59 & -0.14 & 1.11 \\
Epoch 5 & 0.15 & 1.12 & 0.23 & 4.11 & -0.13 & 1.10
\enddata
\end{deluxetable}


Figure \ref{fig:orbdemo} illustrates the results of our orbit retrieval for five epochs of a planet at 20~pc with a 90-day cadence. This particular planet goes undetected in the second epoch, yet we still see an increased similarity between the retrieved and true orbits. In this example, the best-fit orbit after each epoch is the Markov chain link with the maximum posterior probability.

\subsection{Semi-major axis retrieval under different eccentricity scenarios}

The three panels in figures \ref{fig:error} and \ref{fig:precision} show the accuracy and precision---respectively---of semi-major axis retrieval, for different prior assumptions and underlying distributions of orbital eccentricity. The distributions are either both beta (realistic), both uniform (pessimistic), or beta-distributed in the prior and uniform in the underlying truths (i.e., a pathological scenario which may underestimate error).

Figure \ref{fig:error} shows the accuracy of the retrieved $a$ estimates as a function of epochs for a sample of \replaced{50}{100} planets. Errors are retrieved from semi-analytic posteriors of $a$ for the underconstrained first two epochs, while MCMC is used for the third epoch and up. "Accuracy" is the difference between the median of the posterior and the true semi-major axis. 

Figure \ref{fig:precision} is complementary to figure \ref{fig:error} and presents the precisions on the retrieved semi-major axes, along with the sample median of these precisions. Again, MCMC is used for the third epoch on. "Precision" is the half-width of the 1$\sigma$ confidence interval, and it shrinks with each additional observation. This is as expected by degrees of freedom: the inflection point near the third epoch indicates that here we begin to gain less precision with additional measurements.

Our baseline case is shown in the left panel of figures \ref{fig:error} and \ref{fig:precision}. Here, by the second epoch we achieve quite good precision. At $1\sigma$ confidence, we can constrain $a$ to \textless25\% with two epochs, and to $\lesssim$10\% with three epochs. \added{These represent "average" results across planets with true semi-major axes between 0.95 and 1.70~AU.}

The central subplots in figures \ref{fig:error} and \ref{fig:precision} demonstrate a case where our precision is overstated; the prior eccentricity distribution is narrower than the underlying distribution. As shown in Table \ref{tab:zscores}, the standard deviation of the z-scores for this scenario of amiss "mixed" distributions is much greater than unity, which confirms the overstatement.{\footnote{The z-score of a fit, defined here as the ratio between our accuracy and precision, has a standard deviation of about unity if the fit precision reliably encompasses the true value, and the retrieved values are Gaussian.} Therefore, if we were to assume a narrower eccentricity distribution than nature provides, we would not be able to believe our retrieved semi-major axis precisions.

The other endpoint draws eccentricity guesses from a uniform distribution (right panel of figures \ref{fig:error} and \ref{fig:precision}). Despite this pessimistic prior, we can still constrain the semi-major axis to 25\% by the third epoch, and by the fifth epoch, retrieval accuracies are not much different from the best-case scenario. \added{In practice, we would start with a uniform prior until we know better, but the left panel is realistic in the steady-state.}


\subsection{Optimal cadence}

\begin{figure}
\plotone{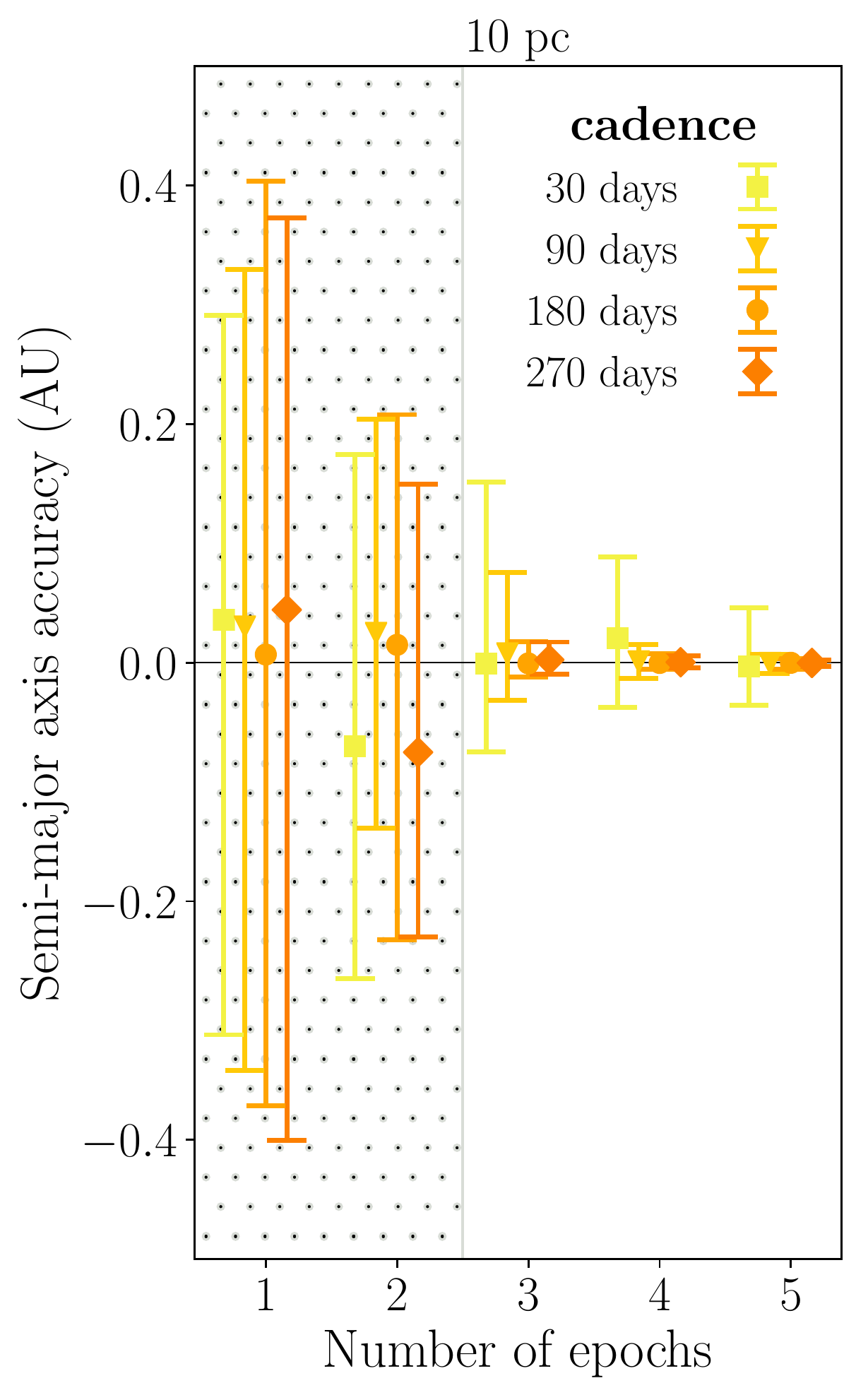}
\caption{\textbf{Accuracy as a function of cadence.} 1$\sigma$ constraints on Markov chain Monte Carlo retrieval accuracies of semi-major axis, as a function of number of epochs, compared across increasing observational cadences. Accuracy is given as $a_{\rm fitted} - a_{\rm true}$, where $a_{\rm fitted}$ is the median of the posterior semi-major axis distribution from MCMC. The error bars show the 16th and 84th percentiles of sample accuracy (\replaced{35}{100} in each scenario), and symbols mark the medians, with colours representing different cadences.
\label{fig:cadence-acc}}
\end{figure}

\begin{figure}
\plotone{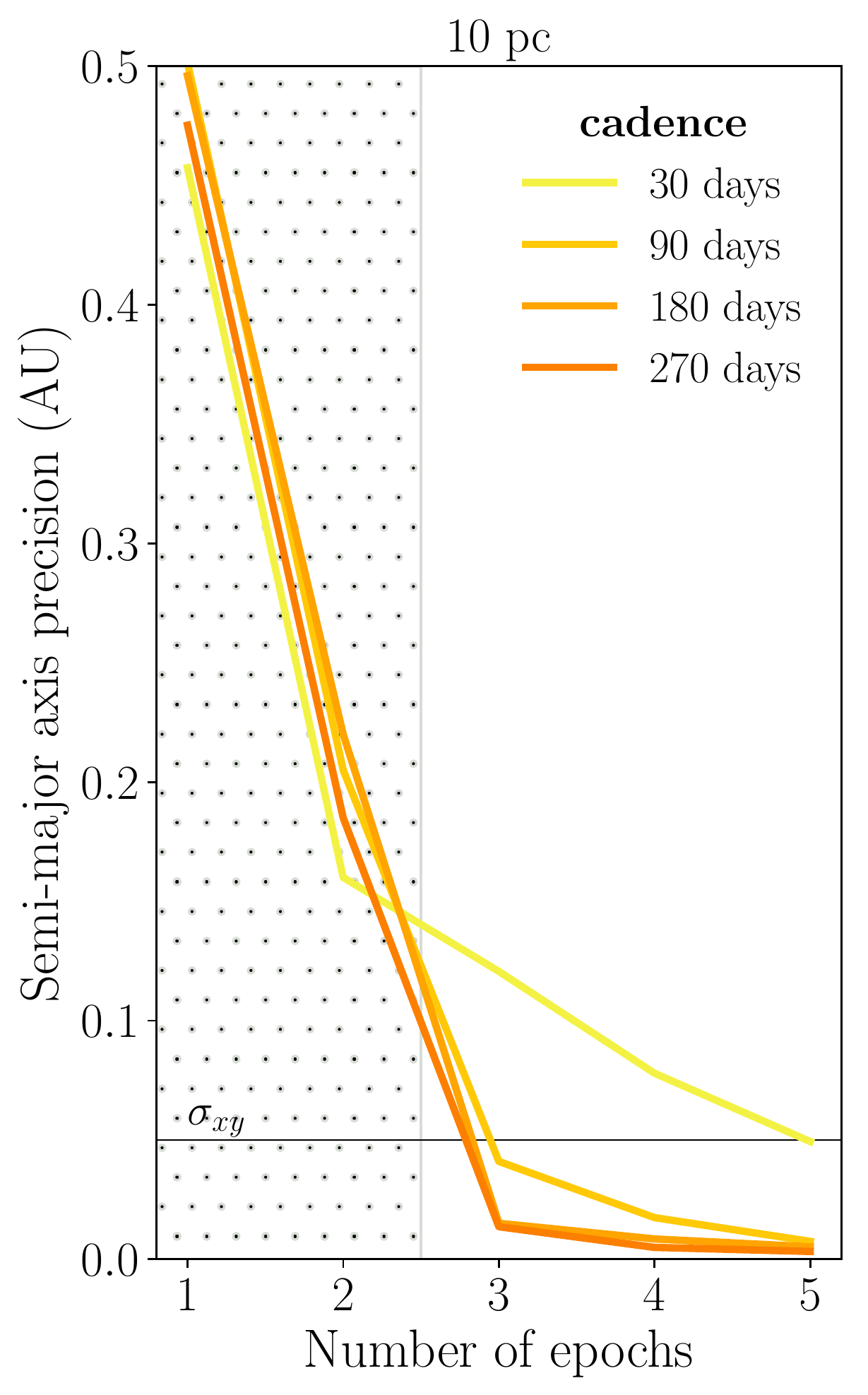}
\caption{\textbf{Precision as a function of cadence.} Sample mean Markov chain Monte Carlo retrieval precisions of semi-major axis, as a function of number of epochs, compared across observational cadences. Precision is given as half the distance between the 16th and 84th percentiles of the retrieved posteriors. Coloured lines show the median sample precision (\replaced{35}{100} in each scenario) for different cadences. The $\sigma_{xy}$ line marks the measurement precision for 3.5~mas astrometry at 10~pc.
\label{fig:cadence-prec}}
\end{figure}

\begin{figure}
\epsscale{1.2}
\plotone{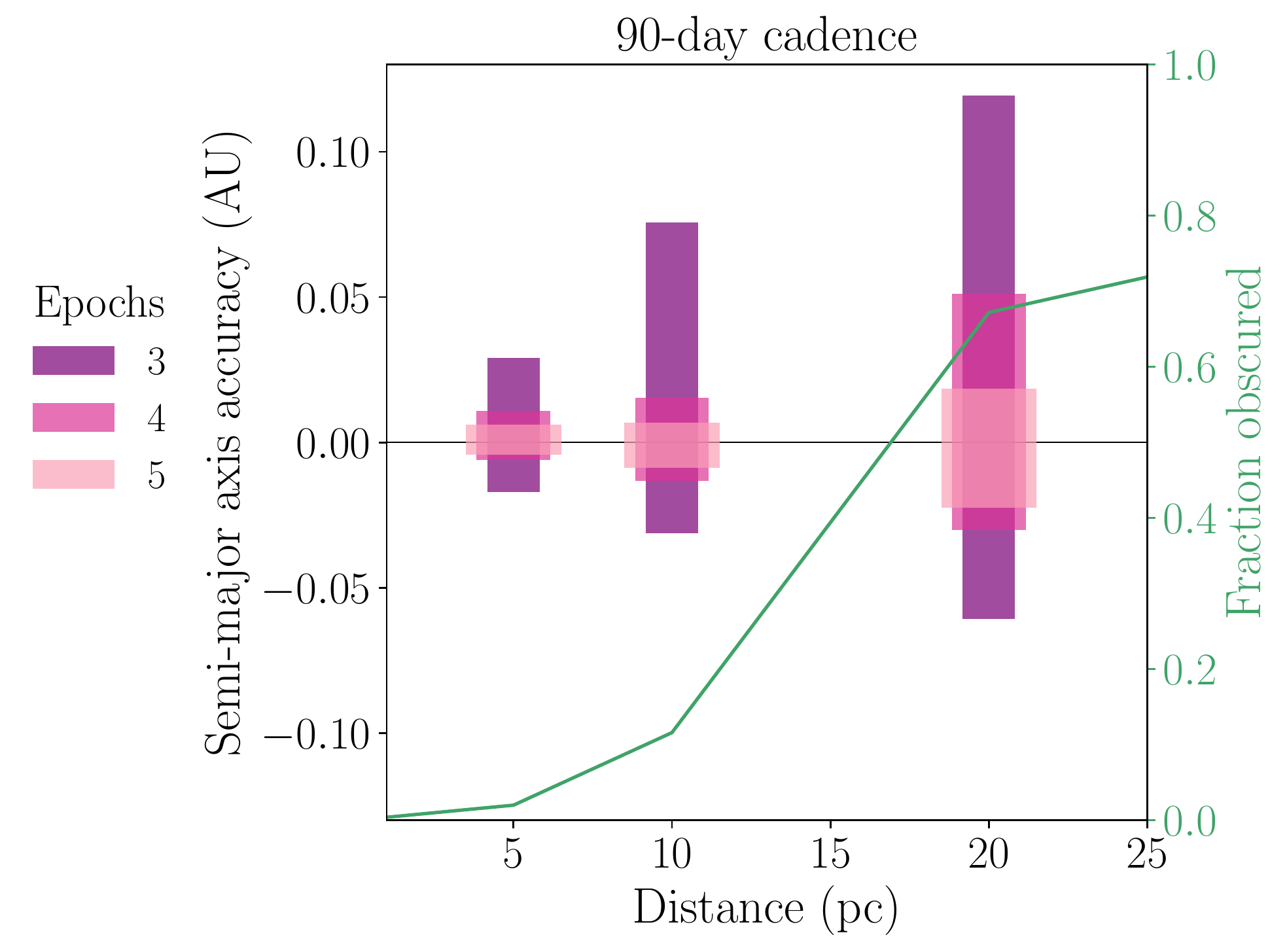}
\caption{\textbf{Targets at distance.} The height of each rectangle represents the 1$\sigma$ confidence interval of the semi-major axis retrieval accuracy for \replaced{50}{100} planets, with shading according to epoch---darker colours are less visits. The green line indicates the actual sample fraction of observations where $\theta_{\rm proj}<\theta_{\rm IWA}$. We would expect accuracy to worsen linearly with distance if only astrometric error mattered (a geometric effect); it worsens faster than linearly because nondetections start to dominate after around 17~pc. \label{fig:distance}}
\end{figure}

\begin{figure*}
\epsscale{1}
\gridline{
\fig{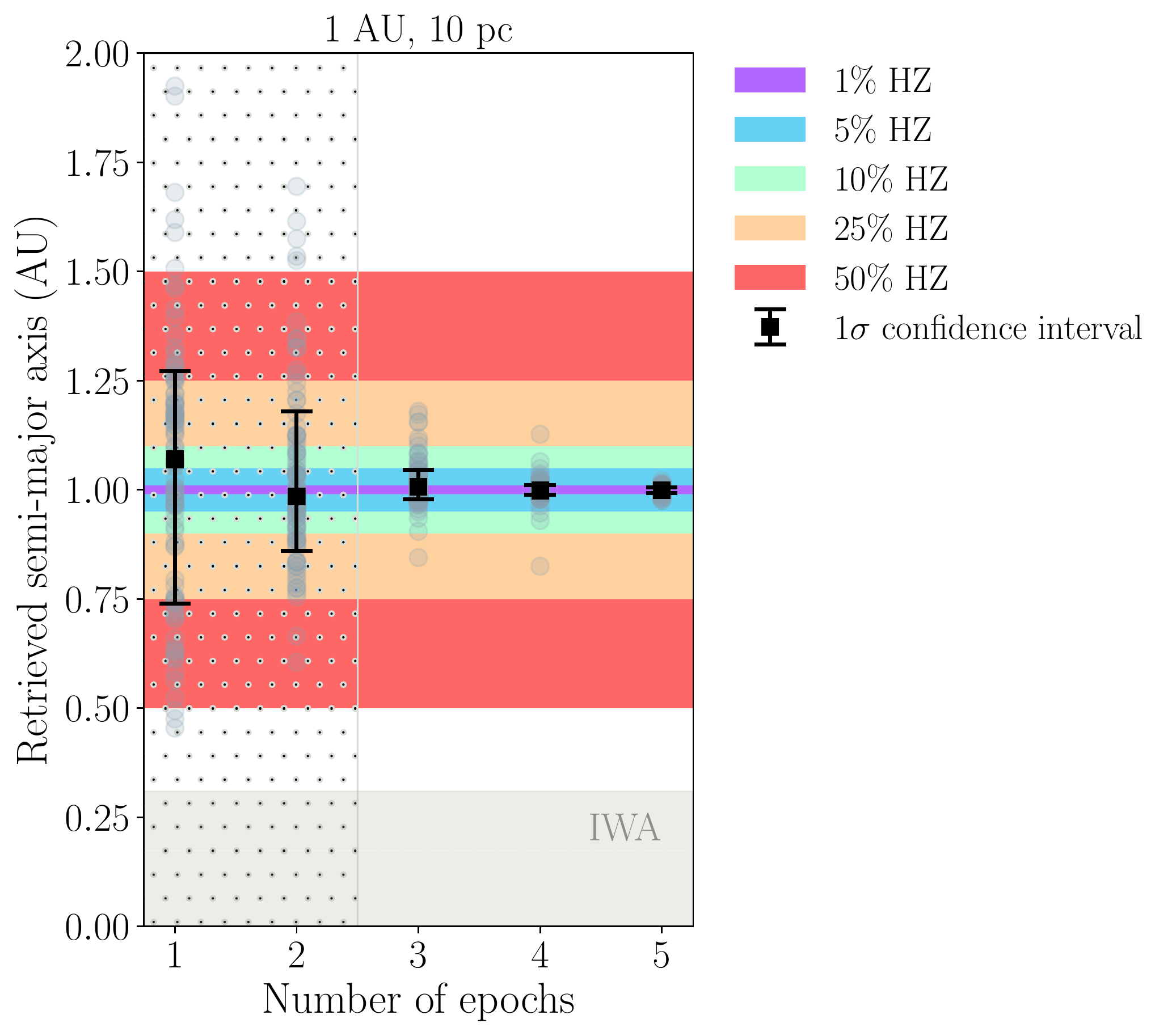}{0.5\textwidth}{}
\fig{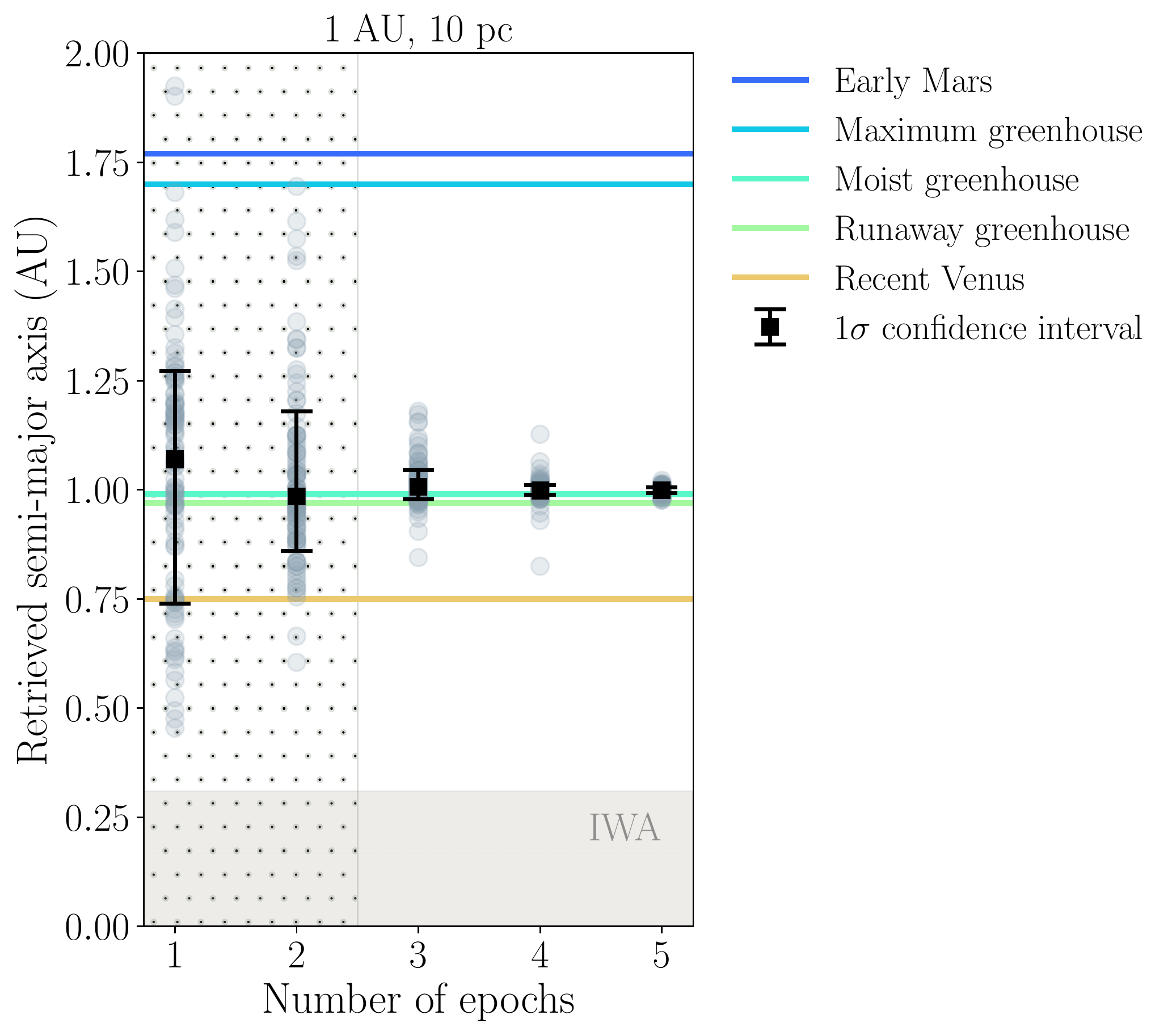}{0.5\textwidth}{}
}
\caption{\textbf{Constraints on semi-major axis compared to habitable zones.} Sample error of retrieved semi-major axes\added{ for a sample of \replaced{50}{100}~planets}, where all planets are assigned a true semi-major axis of 1 AU. The grey hatched area shows the region obscured by the coronagraph inner working angle (31~mas) for a target at 10~pc. Different conceptual habitable zones are superimposed. {\textit{(Left)}:} Coloured swaths represent arbitrary habitable zones relative to the true semi-major axis. For a hypothetical habitable zone with 5\% width, three epochs can constrain an orbit to lie within, at 1$\sigma$. \textit{(Right):} Coloured horizontal lines indicate habitable zone limits from \citet{Kopparapu13}. At 1$\sigma$, constraining a planet to lie within the wider Recent Venus inner limit requires about one epoch.  
\label{fig:HZ}}
\end{figure*}

We ran our experiment for different observational cadences (30, 90, 180 and 270 days) to see which retrieves the most accurate orbits in the least number of epochs. Here, host stars are all at 10~pc and planets have beta-distributed eccentricities. Our "optimal cadence" pertains to planets with periods in or near the G2V habitable zone.

Results are shown in figures \ref{fig:cadence-acc} and \ref{fig:cadence-prec}---by the third epoch, $a$ can be constrained to within 10\% of its true value for 68\% of samples only for cadences 90 days or greater. In half a period, the planet will move to the furthest possible position from its original location. Revisiting the planet at its antipodal point means that we are less likely to miss the planet inside the IWA (e.g., compared to revisiting a quarter-period later), although this is tied to us ensuring the first epoch is a detection.

180- and 270-day cadences result in the best accuracy on semi-major axis at epoch 3, but 180-day cadences for near-360-day orbits may not break formal degeneracies between $\omega_p$ and $\Omega$. Hence the optimal cadence is not necessarily the same for $a$ as for other orbital parameters. While the 270-day cadence also performs well, in practice this long wait could risk losing track of the planet, as discussed below.

\subsection{Effects of distance and astrometric precision}

We have adopted a distance of 10~pc and a measurement error of $\sigma_{\theta}=3.5$~mas for our numerical experiments thus far. However, the number of epochs needed to constrain an orbit is distance-dependent. More-distant targets impede orbit retrievals in two ways: the corresponding measurement error on planet position increases (by a factor of four from 5\textendash20~pc); and the larger projected separation of the IWA is more likely to obscure targets (nearly 0\% nondetections at 5~pc to $\sim$70\% at 20~pc). 

To quantify the effect of distance, we repeated the retrieval experiments for the same planets at 5~pc and 20~pc, assuming 90-day cadences. At our assumed IWA and range of true semi-major axes, HZ planets will be obscured 100\% of the time by 55~pc. This experiment is equivalent to increasing the IWA and astrometric uncertainty. Particular sources of astrometric error and their impact on orbit retrievals are discussed in \citet{Pueyo15}.

Figure \ref{fig:distance} illustrates the change in semi-major axis retrieval accuracy with distance. As expected, the width of the 1$\sigma$ confidence interval increases for more distant targets, for all epochs. 

The encouraging performance of the fit at 5~pc is attributed to the small astrometric error. For these nearest targets we would be capable of measuring $a$ to within \textless2\% by the fourth epoch.

We are only showing this distance dependence of accuracy for one cadence. The optimal cadence is distance-dependent; with increasing distance, obscuration dominates measurement error in worsening the accuracy, preferring 180-day (antipodal) cadences.

\section{Discussion}

The minimum number of epochs we need depends on the science question we are asking.

\subsection{Placing a the planet in the habitable zone}

We first consider hypothetical HZs defined only by how narrow they are. Compare the widths of these hypothetical HZs with the semi-major axis retrievals of planets on 1-AU orbits (figure \ref{fig:HZ}, left panel): at 1$\sigma$ confidence, for an optimistic, 25\%-wide HZ centered on 1~AU, a single epoch of direct imaging generally suffices to nail the semi-major axis accurately. For a pessimistic 1\%-wide HZ, four or five epochs are needed. 

To compare these results with theoretical HZs from the literature: the right panel in figure \ref{fig:HZ} shows the same semi-major axis constraints, to scale with both a LUVOIR-esque inner working angle and with the theoretical HZ inner limits for solar twins from \citet{Kopparapu13}. By the fourth epoch we are 68\% sure a planet is beyond the Moist Greenhouse inner limit. To constrain an orbit beyond the more optimistic Recent Venus inner limit at 0.75 AU at the same confidence, about one epoch is required. 

We are implicitly assuming that the HZ is a function of semi-major axis only. While high eccentricities may negatively affect the stability of surface water \citep{Bolmont16}, the high thermal inertia of extant oceans can buffer against transitions to "snowball" states at apoastron \citep{Dressing10}. Either way, the long-term stability of a planetary climate depends, to the first order, on the average incident flux over the entire orbit \citep{Williams02}, which is proportional to $(1-e^2)^{-1/2}$.






\subsection{Predicting a planet's future location}

In some cases, we would need to constrain all six orbital elements. For example, knowing the entire orbit would help us predict the exoplanetary ephemerides. \replaced{In practice, the accuracy required is set by the detector pixel scale}{The best theoretical precision on extrapolating a planet's position is set by the detector pixel scale, although in practice, the precision we achieve depends on the post-processing}. Using three epochs, we have predicted the position of a planet at the fourth epoch to within a given number of pixels on average. For 180-day cadences, we find the position to be constrained to within one pixel; for 90-day cadences, within two pixels, with five epochs necessary to get down to a one-pixel accuracy. The trade-off is that waiting longer between observations provides better accuracies for the Keplerian parameters, but any errors are amplified by the greater distance the planet will have travelled.
 
On a related note, we have only considered single-planet systems in our study. However, we expect a substantial fraction of planets to occur in multi-planet systems \citep{Zhu18}. This presents an interesting tension with respect to the optimal cadence: for multi-planet systems, we may prefer shorter observational cadences because the planet will move less, so there is less risk of confusing it with other planets. Nevertheless, planet disambiguation in multi-planet systems will be made a bit easier by the fact that we can expect these targets to have near-circular orbits \citep{Eylen18}.

\added{\subsection{Extension to FGK stars}
We have only considered solar-twin stars in this work. However, LUVOIR and HabEx will target a larger range of host stars, whose HZs will scale with stellar mass according to $r/1 \; {\rm AU} \propto \left(M_* / M_{\Sun}\right)^2$. This matters because tighter HZs will be obscured more often inside the IWA. In terms of angular separation, then, this scaling is equivalent to varying the distance to the host star. For example, a planet receiving the same stellar flux as Earth and orbiting a K2V star (0.70~$M_{\Sun}$) will have a separation of 0.49~AU, which is essentially isomorphic to doubling the distance to an equivalent G2V system, as in figure \ref{fig:distance} (barring the astrometric precision change). The optimal cadence for such a system would also scale with orbital period as $P/365 \; {\rm days} \propto \left(M_* / M_{\Sun}\right)^3$, holding to the logic that 90~days $\approx 0.25P$ for solar twins.}

\added{\subsection{Uncertainties on stellar distance}
This model has not considered imperfect knowledge of the distance to the host star. The uncertainty on this distance would propagate linearly to an uncertainty on the planet $(x,y)$ position, which compounds the astrometric measurement error, $\sigma_{\theta}$. LUVOIR and HabEx target stars would be chosen from the Hipparcos and Gaia catalogues \citep{LUVOIR, HabEx}, with known trigonometric parallaxes. For a G2V star within 50~pc (the maximum LUVOIR survey distance), the end-of-mission Gaia parallax uncertainty will be on the order of 0.1~mas \citep{deBruijne2014}; i.e., at most $\pm$1~pc in distance, and an order of magnitude lower than our adopted value of $\sigma_{\theta}$.}

\added{\subsection{Limitations of using astrometry alone} \label{sec:limitations}

This work demonstrates a first step in the orbit determination problem, considering only the planet position data. Yet direct imaging would also measure the brightness of the planet relative to its star. Including this photometric information could both help and hurt the conclusions in this paper. 

On the one hand, the brightness at a given $(x,y)$ position represents a constraint on the orbital phase of the planet. Although the phase functions of our target planets are unknown \textit{a priori}, Lambertian reflection is a fair assumption at phase angles smaller than crescent phase \citep{Robinson2010}.

On the other hand, we have so far ignored that a planet may go undetected at a given epoch not because it is inside the IWA, but because it is imaged at an orbital phase not bright enough for the contrast floor. This would limit our ability to use non-detections as a constraint on the planet position for nearby targets. For more distant targets, the IWA will tend to consume crescent-phase planets either way---we will have less leverage on orbital phase constraints from brightness variations, and missing planets because they are too faint will stop mattering.

Further, photometry brings with it other sources of noise beyond astrometric precision, namely, distinguishing real planets from speckles. This work implicitly assumes that a given contrast floor \citep[e.g., $10^{-10}$;][]{LUVOIR} already accounts for the removal of speckles in post-processing. Including photometry in our model may increase the number of visits to achieve the same precision, although it would not increase the required number of detections.
}

\smallskip 

The question this work has asked is: how well can one locate a planet in the habitable zone? Under the assumption of one planet per star, we find that 180- and 270-day cadences have the best precision and accuracy. In a realistic mission, however, these long cadences would likely be poor at resolving confusion between multiple planets in the same system. A 90-day cadence is a good compromise.

Even if the third image placed a planet in the habitable zone at 95\% confidence, we would still need more epochs to establish its orbit before pursuing expensive spectroscopy. Yet the point is that we could begin prioritizing targets after only one or two epochs. This quantifies a key parameter in Design Reference Missions for future direct imaging concept missions HabEx and LUVOIR. In the LUVOIR study, constraining orbits within the habitable zone is the third step to identifying a habitable planet, after (1) establishing the target star list and (2) performing multi-colour point source photometry to rule out background objects \citep{LUVOIR}. Our work finds that the orbit-constraining step can be done more efficiently than before, reducing the minimum number of observations from six to three. This extra time could be spent characterizing more planets.

\acknowledgments

This work is supported by the McGill Space Institute, the TEPS training program, and an NSERC Discovery Grant. The authors wish to thank the HabEx AEIWG for their useful conversation: especially Leslie Rogers, Eric Nielsen, and Scott Gaudi. Comments from an anonymous reviewer also greatly improved the quality of this manuscript. All research was conducted on the territory of the Kanien'keh\'a:ka, the keepers of the Eastern Door of the Haudenosaunee Confederacy. 

\bibliography{ms}

\appendix
\section{Forward model}\label{sec:forward}

We assume that all planets are on bound Keplerian orbits.

If we hover with the planet's orbital plane below us and our right ears towards the +$x$ reference direction, we will "see" an ellipse. Its shape can be described parametrically as the first two dimensions of a three-dimensional matrix:
\begin{equation}\label{eq:xy}
\begin{bmatrix}
x_k\\y_k\\z_k
\end{bmatrix}
= \begin{bmatrix}
  a\,\cos(M_k) - c\\
  b\,\sin(M_k)\\
  0
 \end{bmatrix},
 \end{equation}
where $a$ is the semi-major axis, $b = a\sqrt{1-e^2}$ is the semi-minor axis, $c=\sqrt{a^2 - b^2}$ is an $x$-intercept, $M_k$ is the mean anomaly at the $k$th epoch, and the orbital plane is defined by $\bm{x}$-$\bm{y}$.

The ellipse we see in the detector plane, $\bm{x^\prime}$-$\bm{y^\prime}$, has been rotated through three angles in the order $\Omega$, $i$, $\omega_p$. There is a Euclidean rotation matrix corresponding to this series of rotations:


\begin{equation}
\begin{bmatrix}
x^\prime \\ y^\prime \\ z^\prime
\end{bmatrix} =
\begin{bmatrix}
\cos\gamma & -\sin\gamma & 0\\
\sin\gamma & \cos\gamma & 0\\
0 & 0 & 1
\end{bmatrix}\cdot
\begin{bmatrix}
1 & 0 & 0 \\
0 & \cos\beta & -\sin\beta \\
0 & \sin\beta & \cos\beta
\end{bmatrix}
\cdot
\begin{bmatrix}
\cos\alpha & -\sin\alpha & 0 \\
\sin\alpha & \cos\alpha & 0\\ 
0 & 0 & 1
\end{bmatrix}
\cdot
\begin{bmatrix}
x \\ y \\ z
\end{bmatrix},\label{eq:rotation}
\end{equation}
where $\alpha=\Omega$, $\beta=i$, and $\gamma=\omega_p$. 

The projected separation in the image plane is the Euclidean distance, $\sqrt{(x^\prime)^2+ (y^\prime)^2}$.

\subsection{Planets move}

Both the cadence, $\delta t$, and the walker's current "guesses" of $a$ and $M_0$ control the planet's position at the $k$th epoch (i.e., set by the instantaneous mean anomaly $M_k$):
\begin{equation}\label{eq:xi}
M_k = M_0 + \sqrt{\frac{\mu}{a^3}}\; k\delta t \; \frac{86400 \;\rm{sec}}{\rm{day}},
\end{equation}
where $\sqrt{{\mu}/{a^3}}$ is the mean motion of the planet (in radians per second) and $\mu$ is the standard gravitational parameter.

\end{document}